\documentstyle[prl,aps]{revtex}

\begin{document}
\draft
\twocolumn[\hsize\textwidth\columnwidth\hsize\csname 
@twocolumnfalse\endcsname
\title{Vortex configurations, matching, and domain structure in large 
arrays of artificial pinning centers }
\author{S. B. Field,$^1$ S. S. James,$^{1,}$\cite{SebAddress} J. Barentine,$^1$\\
V. Metlushko,$^{2}$ G. Crabtree,$^{2}$
H. Shtrikman,$^{3}$
B. Ilic,$^{4}$ and
S. R. J. Brueck$^{5}$
}
\address{
$^1$Department of Physics, Colorado State University,
Fort Collins CO 80523\\
$^2$MSD, Argonne National Laboratory, Argonne, IL 60439-4845\\
$^3$Weizmann Institute of Science, Rehovot, Israel\\
$^{4}$School of Applied and Engineering Physics, Cornell University, Ithaca, 
NY 14853\\
$^{5}$Center for High Technology Materials, University of New
Mexico, Albuquerque, NM 87131
}

\date{\today}
\maketitle
\begin{abstract}
High-resolution scanning Hall probe microscopy has been used to image 
vortex configurations in very large periodic arrays of artificial 
pinning sites.  Strong matching effects are seen at fields where 
either one or two vortices can sit at a site; with three vortices per 
site, however, no clear matching is observed.  Matching effects 
have been also been observed at several fractional multiples of the 
matching field, including 1/5, 1/4, 1/3, 1/2, and 3/4.  These 
fractional values are characterized by striking domain structure and 
grain boundaries.
\end{abstract}

\narrowtext
\vskip0pc]

%---------------------- Introduction ------------------------

The behavior of superconducting vortices in the presence of a periodic 
array of holes reveals rich and unexpected static and dynamic 
phenomena.  Transport \cite{Fiory78,VanLook99,Puig98} and 
magnetization 
\cite{Baert95,Moshchalkov96,Moshchalkov98,Metlushko99,Metlushko99a} 
studies show distinctive features at matching fields where the 
vortex structure is commensurate with the hole array.  
Commensurability can arise from vortex configurations 
\cite{Baert95,Moshchalkov96,Moshchalkov98} in which each hole contains 
an integer number of flux quanta, or from interstitial structures 
where vortices occupy the regions between flux-saturated holes.
Interesting configurations also arise at fractional matching fields, 
where the occupation number of each hole or the number of 
interstitial vortices forms a superstructure locked to the basic hole 
array \cite{Baert95,Metlushko99a}.  

Although magnetization and transport studies have elucidated much of 
the basic phenomenology of the array/vortex system, they can measure 
only its {\it global} properties and cannot deduce details of the {\it 
local configurational} vortex state.  More recently Lorentz microscopy 
has been used to image vortex configurations \cite{Harada96}, but only 
in relatively small arrays consisting of $13 \times 13 = 169$ pinning 
sites.  In this paper we present large-area scanning Hall probe 
microscope (SHM) images of vortex configurations in arrays containing 
$\approx 10^{6}$ holes.  Our images span some 5000 holes, and so 
yield important information on the {\em large-scale} structure of the 
vortex configurations.  These studies reveal striking multi-quantum 
and interstitial vortex patterns in a square-periodic hole array.  At 
fractional matching fields we resolve distinctive domains 
of phase-slip related vortex superstructures that are separated by 
domain walls with characteristic internal structure.

%-------------------- Experimental Details ----------------------

The SHM uses a large-range scanner yielding 
$\approx$ 130 $\times$ 130 $\mu$m images.  The sample investigated was 
a 100-nm-thick niobium film with 0.3-$\mu$m-diameter holes on a square 
lattice.  The lattice constant $a$ of the film was 1.870 $\mu$m, 
yielding an expected first matching field $H_{m}=\Phi_{0}/a^{2}$ = 
5.913 G. The array was produced using a lithographic technique based 
on the interference of light \cite{Metlushko99}, which yields very 
large and uniform arrays (here $\approx$ 2 mm $\times$ 3 mm).  Near 
$T_{c} = 8.37$ K, the array exhibits clear cusps in its magnetization 
at $H_{m}$ and $2 H_{m}$, but not at any higher integral multiples of 
$H_{m}$.

%------------------------ Overview ------------------------

SHM images near several matching fields are shown in Fig.~1.  For each 
image, the samples were field-cooled from above $T_{c}$ to a base 
temperature below 3 K where the images were taken.  The top row of 
Fig.~1 shows the progression in the vortex configurations as the 
applied field $H$ is varied around $\overline{H}\equiv H/H_{m} =$ 0.  
At $\overline{H} \approx 0$ we see several isolated vortices (dark 
spots).  The actual vortex diameter is much smaller than these spots, 
whose size is determined by the Hall probe resolution.  A somewhat 
darker spot, just to the right and below center, appears in all scans 
and is presumably associated with a physical defect in the hole 
lattice, which allows extra flux to sit at that point.  As the field 
is increased from zero, vortices begin to enter the sample in larger 
numbers, as seen in the $\overline{H} = 0.075$ and 0.15 images. A similar 
progression is observed for {\em negative} field increments away from 
zero, as shown at $\overline{H} = -0.075$ and $-0.15$.  Here the 
vortices appear as light spots, indicating that they are oppositely 
directed to those seen in positive fields.

As we continue to increase the field towards what we will be able to 
identify as the first matching field $H_{m}$ ($\overline{H} = 1$), we 
begin to see a very remarkable progression (Fig.~1, second row).  For 
instance, at an applied field of $\overline{H} = 0.85 = 1 - 0.15$, the 
image looks statistically identical to the image at $\overline{H} = 
-0.15$!  And, remarkably, the image at $\overline{H} = 1$ is 
essentially indistinguishable from that at $\overline{H} = 0$.  It is 
important to note here that in each image the average value of the 
field has been subtracted out.  This is why the $\overline{H} = 1$ 
image at about 6 G and the $\overline{H} = 0$ image appear to have 
about the same overall gray level.  We can then understand the appearance 
of the $\overline{H} = 1$ scan as follows.  The field increment from 
an exactly $H = 0$ image (no spots) to the next no-spot image is 
5.929~G, very close to the first matching field of 5.913~G deduced 
from the lattice geometry.  Thus we identify this later spot-free 
image at 5.929~G with the first matching field or $\overline{H} \equiv 
1$.  The smooth gray background near $\overline{H} \approx 1$ can then 
be explained in the following way.  Exactly at $\overline{H} = 1$, each hole 
contains exactly one vortex, leading to a dense and uniform 
configuration of vortices.  Evidently the spatial resolution of our 
Hall probe is not good enough to image individual vortices when they 
are this close together {\em and perfectly ordered} 
\cite{Davidovic96}, so this regular array of vortices appears as a 
smooth gray background.  The few black spots in the $\overline{H} 
\approx 1$ image are thus (easily visible) {\it extra} vortices.

By the same token the white spots in the images at $\overline{H} = 
0.85$ and 0.925 are not antivortices, as they were in the $-0.15$ and 
$-0.075$ images.  Instead, they are {\em vacancies} in a smooth 
background of filled holes.  Since vacancies---holes with {\em no} 
vortex---have a lower (whiter) field than the broad gray areas which 
are filled with vortices, they appear as white dots.

The third row of Fig.~1 shows the vortex configurations as the field 
is further increased through the second matching field ($\overline{H} 
= 2$).  A very similar progression is observed.  At $\overline{H} 
\approx 2$, there is again a smooth background populated by a few 
black spots.  Again, we interpret this background as the highly 
ordered state with now {\em two} vortices sitting in each hole; black 
spots for $\overline{H} > 2$ are again extra vortices, and white spots 
for $\overline{H} < 2$ now represent a hole with only {\em one} 
vortex.  At the {\em third} matching field ($\overline{H} = 3$), 
however, the appearance is radically different (Fig.~1, bottom right).  
Instead of the smooth gray background we would expect if three 
vortices sat in each hole, we see a rather muddled image with no 
discernible structure.  We believe this different appearance results 
from the sudden presence of {\em interstitial} vortices above 
$\overline{H} = 2$.

To probe this issue more directly, we have taken closeup scans just 
below and above $\overline{H} = 2$.  Figure~2(a) shows a 25~$\mu$m 
$\times$ 28~$\mu$m scan at a field $\overline{H} = 2 - 0.084$, and 
Fig.~2(b) shows $\overline{H} = 2 + 0.084$.  Also shown is the 
least-squares-fit positions of the holes as determined from an image 
taken at $\overline{H} = 1/2$ where the hole positions are unambiguous 
[see, e.g., Fig.~3(b)].  We also have used an absolute position sensor 
\cite{SensorPaper} mounted on the scanner head to compensate for any 
possible drifts between images.  In Fig.~2(a) each white spot (i.e., a 
``missing'' vortex) clearly sits directly on a hole, indicating as 
well that all {\em vortices} also sit on the holes.  However, in 
Fig.~2(b) the black spots, which are now extra vortices, sit variously 
on holes {\em or} on interstitial sites.  As is well known, each hole 
in a film can hold up to a {\it saturation number} $n_{s}$ of vortices 
\cite{Mkrtchyan72}; with our material parameters we estimate that 
$n_{s}$ is near the boundary between 2 and 3.  With the 
sputtering/lift-off lithography used here \cite{Metlushko99} there 
will be inevitable fluctuations in the hole diameter, perhaps yielding 
$n_{s}$ = 2 for some holes and $n_{s}$ = 3 for others.  Vortices 
nucleating near $n_{s}$ = 2 holes would end up as interstitials, while 
those near $n_{s}$ = 3 sites could sit in a hole.  This 
random-appearing admixture of vortices on holes and in interstitial 
sites would, by $\overline{H} = 3$, lead to the very disordered 
configuration observed at that field (Fig.~1).

%---------------------- Submatching ------------------------

It is also possible to have matching effects at 
non-integer multiples of $H_{m}$.  Indeed, distinct features at 
$\overline{H} = 1/4$, 1/2, and possibly 1/5 (or 3/16), 1/8, and 1/16 
have been observed in magnetization studies 
\cite{Baert95,Metlushko94}, and Harada {\it et al.}\ \cite{Harada96} 
have imaged non-integer matching at 1/4, 1/2, 3/2, and 5/2.  In our 
imaging experiments, we also observe matching effects at 
several of these fractional fields, as well as some not previously 
observed.  We find that vortex configurations at fractional matching 
fields are characterized by striking domain structure and associated 
grain boundaries.

By far the strongest submatching effects occur at $\overline{H} = 
1/2$.  Figure~3(a) shows that the configuration consists of large 
areas of well-matched vortices separated by curious stripe-like 
boundaries.  The closeup in Fig.~3(b) reveals that the smooth regions 
consist of vortices occupying every other hole in a checkerboard-like 
fashion.  Regions of different ``polarity''---with the vortices 
occupying either the ``red'' or ``black'' squares of the 
checkerboard---are separated by striped grain boundaries 
\cite{Reichhardt96}.  Fig.~3(c) schematically shows how a polarity 
shift results in rows or columns of alternating {\em pairs} of 
vortices and vacancies, which appear in the images as the striking 
boundary features.  At the junctions between north-south and east-west 
boundaries there is always a bright or dark spot, whose origin is 
again clear from Fig.~3(c).  From considerations of the boundaries, we 
have mapped out the domains of differing polarity [inset, Fig.~3(a)].  
One curious feature of these grain boundaries is that they run 
predominately north-south, indicating a breaking of the square 
symmetry of the lattice in these two directions.

Weaker but still striking matching effects occur at $\overline{H} = 
1/4$ as well as its {\em complimentary} \cite{CompNote} field of 
$\overline{H} = 1 - 1/4 = 3/4$ (Fig.~4).  We note immediately that the 
matching here is much less distinctive than at 1/2, 1, or 2.  
Nonetheless, there do exist some relatively large ordered domains.  
Zooming in on the boundary of such a patch [outlined in Figs.~4(a) and 
4(b)] allows us to explore the vortex configuration in detail [Figs.  
4(c) and 4(d)].  Here we can clearly see areas with well-ordered 
vortices interspersed with phase slips (diagonal light and dark 
segments) and other, more complex features.  The vortex configuration 
in the ordered regions is shown to scale schematically in Fig.~4.  
Notice that the configuration may be viewed in terms of alternating 
empty and half-filled rows \cite{Harada96,Watson97}; this again 
indicates the presence of a symmetry-breaking field which selects this 
particular grain orientation.  One possible source of this asymmetry 
could be a small difference in the interhole spacing $a$ in the 
horizontal and vertical directions.  However, no such asymmetry is 
visible at the $\approx$ 0.1\% accuracy of our optical diffraction 
measurements of $a$.

%% 
 % Little in the way of	a general theory has been developed	regarding 
 % vortex configurational groundstates in periodic arrays of pinning 
 % sites.  Baert {\it et al.} \cite{Baert95} claim particularly	stable 
 % configurations exist	when the vortices form square lattices with	sides 
 % larger than $a$.	 More recently Watson \cite{Watson97} has studied the 
 % general problem of the ground-state configurations of repulsive 
 % particles constrained to	a two-dimensional square lattice, a	problem	
 % of obvious relevance	here.  Finally,	several	configurational	
 % groundstates	have been predicted	for	the	related	case of	vortices in	a 
 % superconducting wire	network	\cite{Teitel83,Hallen93}.
 %%

Besides the 1/2, 1/4, and 3/4 states already discussed, we have found 
several other relatively well-ordered states at fractional fillings $0 
\le \overline{H} \le 1$.  Figure~5(a) shows the configuration at 
$\overline{H} = 1/5$.  Although there are no large domains, in much of 
the image we can observe a general tendency towards forming a 
structure which consists of a square lattice of vortices tilted by an 
angle $\arctan(1/2) \approx 26.6^{\circ}$ \cite{Baert95,Watson97}; 
this structure is shown in detail in the inset.  There is also 
striking structure at $\overline{H} = 1/3$ [Fig.~5(b)].  Here, the 
configurations consist of slope-one diagonal stripes of (unresolved) 
vortices separated by two empty diagonal stripes.  Such a 
configuration has been predicted by Watson \cite{Watson97} for the 
general case of repulsive particles on a square lattice, and by Teitel 
and Jayaprakash \cite{Teitel83} for the related case of vortices in a 
superconducting wire network \cite{Hallen93}.  Again, domains with 
this precise structure are small, and there are a large number of 
defects present.  There is very weak ordering present at $\overline{H} 
= 2/5$ [Fig.~5(c)], where Teitel and Jayaprakash \cite{Teitel83} 
predict states consisting of two empty diagonals and then a 
full-empty-full diagonal sequence.  This later sequence has the 
appearance of a checkerboard state, and is barely visible between the 
bright (empty) diagonal stripes of the inset to Fig.\ 5(c).  Finally, 
we show an image taken ``at'' the irrational field $\overline{H} = 
(\sqrt{5} - 1)/2 \approx 0.618$.  No apparent order is present in this 
case.

Vortex configurations in square-periodic hole arrays thus reveal 
remarkably complex patterns reflecting the interplay between the 
pinning energy of the hole array and the interaction energy of the 
multi-quantum and interstitial flux structures.  Our images 
show clear multi-quantum occupation of the holes up to saturation, 
followed by the appearance of interstitial Abrikosov vortices at 
higher fields.  At fractional filling factors, domains of well-ordered 
vortices are separated by a network of domain boundaries displaying 
remarkably simple order for half filling and increasing complexity at 
other fillings.  There appears to be a nontrivial 
set of ordered fractional states, a complete inventory of which awaits 
a general theory correctly incorporating the several competing 
interactions present.  Our Hall probe images thus provide direct visual 
evidence for a rich diversity of collective ground states arising from 
competing commensurability, order, and randomness in the vortex phase 
diagram of periodic hole arrays.

This work was supported by NSF Grant DMR-9701532 (SBF, SSJ, JB) and 
the US Department of Energy, Office of Basic Energy Sciences-Materials 
Sciences, under contract \#W-31-ENG-38 (VM, GWC).

%----------------------- References --------------------------

%----------------------- Figure Captions --------------------------

\begin{figure}
\caption{SHM images obtained at applied magnetic fields $H$ near the 
matching fields $\overline{H} \equiv H/H_{m} =$ 1, 2, and 3.  The scans 
are 124 $\mu$m $\times$ 138 $\mu$m, contain some 5000 holes, and 
have a full scale of 0.73~G. Note the striking similarity between the 
vortex configurations near $\overline{H}$ = 1 and 2 and those near 
$\overline{H}$ = 0.  The smooth backgrounds at $\overline{H}$ = 1 (2) 
consist of holes uniformly filled with 1 (2) vortices.  At 
$\overline{H} = 3$, however, the vortex configuration is highly 
disordered because of weakly-pinned interstitials competing with 
vortices in holes.  The dark black spot is a local defect in the 
sample.}
\label{fig1}
\end{figure} 

\begin{figure}
\caption{Vortex configurations (a) just below ($\overline{H} = 1.916$) 
and (b) just above ($\overline{H} = 2.084$) the second matching field.  
The small dark circles are the positions of the holes.  Below 
$\overline{H} = 2$ the vacancies (white spots) all sit directly on 
holes; thus all vortices must as well.  Above $\overline{H} = 2$, the 
extra vortices (dark spots) sit on both hole {\it and} interstitial 
sites.}
\label{fig2}
\end{figure}

\begin{figure}
\caption{(a) Vortex configuration at $\overline{H} = 1/2$.  There are 
large areas of vortices arranged in a checkerboard pattern; domains 
[outlined in inset to (a)] of opposite checkerboard polarity are 
separated by stripe-like grain boundaries.  A closeup of this 
configuration is shown in (b) and schematically in (c).}
\label{fig3}
\end{figure}

\begin{figure}
\caption{Vortex configurations at $\overline{H} = 1/4$ (a) and $3/4$ 
(b).  Selected regions are expanded in (c) and (d).  The 
configurations consist of relatively small well-ordered regions 
surrounded by complex disordered boundaries.  Also shown are the 
inferred vortex arrangements in the ordered regions.  Light circles 
represent empty holes, and dark circles occupied holes.}
\label{fig4}
\end{figure}

\begin{figure}
\caption{Vortex configurations at several fractional values of 
$\overline{H}$.  Also shown is the ``irrational'' field $\phi = 
(\sqrt{5} - 1)/2$.}
\end{figure}


\begin{references} 

\bibitem[*]{SebAddress}Present address: IRC in Superconductivity,
Madingley Road, Cambridge, CB3 0HE, UK.

\bibitem{Fiory78}A. T. Fiory, A. F. Hebard, and S. Somekh, Appl. Phys. Lett. 32, 73 (1978)
\bibitem{VanLook99}L. Van Look {\it et al.}, Phys. Rev. B 60, 6998, (1999)
\bibitem{Puig98}T. Puig {\it et al.}, Phys. Rev. B 58, 5744 (1998).
\bibitem{Baert95}M. Baert {\em et al.}, Europhys. Lett. {\bf 29}, 157 (1995).
\bibitem{Moshchalkov96}V. V. Moshchalkov {\it et al.}, Phys. Rev. B 54, 7385 (1996).
\bibitem{Moshchalkov98}V. V. Moshchalkov {\it et al.}, Phys. Rev. B 57, 3615 (1998).
\bibitem{Metlushko99}V. Metlushko {\it et al.}, Phys. Rev. B {\bf 59}, 603 (1999).
\bibitem{Metlushko99a}V. Metlushko {\it et al.}, Phys. Rev. B 60, 12585 (1999a).
\bibitem{Harada96}K. Harada {\em et al.}, Science {\bf 274}, 1167 
(1996).
\bibitem{Davidovic96}D. Davidovic {\em et al.}, Phys. Rev. B
{\bf 55}, 6518 (1996).
\bibitem{SensorPaper}S. Field and J. Barentine, Rev. Sci. Instrum. 
(to be published).
\bibitem{Mkrtchyan72}G. S. Mkrtchyan and V. V. Shmidt, Sov. Phys. 
JETP {\bf 34}, 195 (1972).
\bibitem{Metlushko94}V. Metlushko {\it et al.}, Solid State Commun. {\bf 91}, 331 (1994).
\bibitem{Reichhardt96}C. Reichhardt {\it et al.}, Phys. Rev. B {\bf 
54}, 16108 (1996).
\bibitem{CompNote}Indeed, for {\it all} configurations observed at 
fractional field $\overline{H} < 1$, essentially identical 
configurations (with vortices and vacancies exchanged) are observed at 
a field $1 - \overline{H}$.
\bibitem{Watson97}G. I. Watson, Physica A {\bf 246}, 253 (1997).
\bibitem{Teitel83}S. Teitel and C. Jayaprakash, Phys. Rev. B {\bf 
27}, 598 (1983); Phys. Rev. Lett. {\bf 51}, 1999 (1983).
\bibitem{Hallen93}H. D. Hallen {\it et al.}, Phys. Rev. Lett. {\bf 
71}, 3007 (1993).



\end{references}
\end{document}